\documentclass[sigconf,nonacm]{acmart}

\AtBeginDocument{%
  \providecommand\BibTeX{{%
    \normalfont B\kern-0.5em{\scshape i\kern-0.25em b}\kern-0.8em\TeX}}}

\setcopyright{acmcopyright}
\copyrightyear{2022}
\acmYear{2022}
\acmDOI{}

\acmConference[]{}{}{}
\acmPrice{}
\acmISBN{}
\usepackage{caption}
\usepackage{subcaption}
\usepackage{graphicx,array}
\usepackage{soul}
\usepackage{hyperref}

\begin{document}

\title{A White-Box Adversarial Attack Against a Digital Twin}



\author{Wilson Patterson}
\affiliation{%
  \institution{Mississippi State University}
  \country{Mississippi State, MS, USA}
}
\email{wep104@msstate.edu}
\author{Ivan Fernandez}
\affiliation{%
  \institution{Mississippi State University}
  \country{Mississippi State, MS, USA}
}
\email{iaf28@msstate.edu}
\author{Subash Neupane}
\affiliation{%
  \institution{Mississippi State University}
  \country{Mississippi State, MS, USA}
}
\email{sn922@msstate.edu}
\author{Milan Parmar}
\affiliation{%
  \institution{Mississippi State University}
  \country{Mississippi State, MS, USA}
}
\email{parmar@cse.msstate.edu}
\author{Sudip Mittal}
\affiliation{%
  \institution{Mississippi State University}
  \country{Mississippi State, MS, USA}
}
\email{mittal@cse.msstate.edu}
\author{Shahram Rahimi}
\affiliation{%
  \institution{Mississippi State University}
  \country{Mississippi State, MS, USA}
}
\email{rahimi@cse.msstate.edu}

\begin{abstract}
Recent research has shown that Machine Learning/Deep Learning (ML/DL) models are particularly vulnerable to adversarial perturbations, which are small changes made to the input data in order to fool a machine learning classifier. The Digital Twin, which is typically described as consisting of a physical entity, a virtual counterpart, and the data connections in between, is increasingly being investigated as a means of improving the performance of physical entities by leveraging computational techniques, which are enabled by the virtual counterpart. This paper explores the susceptibility of Digital Twin (DT), a virtual model designed to accurately reflect a physical object using ML/DL classifiers that operate as Cyber Physical Systems (CPS), to adversarial attacks. As a proof of concept, we first formulate a DT of a vehicular system using a deep neural network architecture and then utilize it to launch an adversarial attack.
We attack the DT model by perturbing the input to the trained model and show how easily the model can be broken with white-box attacks.

\end{abstract}



\maketitle

\vspace{-2mm}
\section{Introduction \& Background}

The evolution of computing, communication, and sensing technologies have resulted in the implementation of internet-controlled Cyber Physical Systems (CPS). Cyber-attacks have evolved in tandem with the development of new technologies to become more effective against this CPS ecosystem. As demonstrated by the Stuxnet attack \cite{5772960} and the Colonial Pipeline hack \cite{copipe}, attackers can now digitally disrupt physical equipment. These cyber-attacks have the ability to target systems so as to acquire valuable intel, obtain personal information, or steal money. This creates a new area of concern for the safety and security of CPSs, as cyber-threats are now physical and may endanger human lives. 

A Digital Twin (DT) is an important part of the CPS ecosystem. It is defined as an ultra-realistic, ``multi-physics, multi-scale, probabilistic simulation of a vehicle or system that uses the best available physical models, sensor updates, fleet history, etc., to mirror the life of its twin" \cite{shafto2012modeling,banerjee2017generating}. DTs have been implemented in a  variety of fields and play a vital role in the function of many CPSs. The reliance on DT devices for automotive industry, military, and medical functions has increased the potential risk of adverse effects, if these systems become compromised. Medical cyber-physical systems (MCPS) \cite{lee2011challenges} are examples of systems that, if compromised, could allow an attacker to not only access a patient’s data but also attack physical devices used to diagnose, monitor, or control their health. Finlayson et al. \cite{finlayson2019adversarial} have demonstrated that a calculated Adversarial Machine Learning (AML) attack can be used to apply perturbations to highly accurate medical image classifiers, resulting in the misclassification of medical images to occur with a high degree of certainty. Vehicles are another source of physical safety concern. Many modern vehicles are outfitted with sensors that monitor the vehicle's functions or enable self-driving capabilities. AML attacks can target the ML models that drive the decisions of autonomous systems, causing misclassification of road signs or other inputs for traffic conditions \cite{eykholt2018robust}. Efforts must be undertaken to secure DTs to improve trustworthiness of the entire CPS ecosystem. 



Adversarial attacks are classified as either white-box or black-box attacks. In a white-box setting, the attacker has complete knowledge of the ML model's architecture, gradients, and parameters. Whereas in a black-box setting, the attacker might have some knowledge of the ML but no access to the architecture, gradients, or parameters. The fast gradient sign method (FGSM) was introduced by Goodfellow et al. \cite{fgsm} as a white-box attack that aims to misclassify adversarial inputs. By using the model parameter, FGSM attempts to produce an incorrect prediction by calculating the amount of perturbation to add to an input that maximizes the loss function. The Adversarial Robustness Toolbox (ART) \cite{art_paper} is a Python library that implements state-of-the-art attacks and defenses. ART was developed by IBM, which is now hosted by the Linux Foundation AI and Data Foundation. 
ART is framework-independent and can handle all major machine-learning tasks.

In this paper, we will focus on exploiting the integrity of a DT system. We will demonstrate a white-box adversarial attack against our DT system using a Machine Learning (ML) model as a proof-of-concept. By applying calculated perturbations, we will cause confusion between the physical half of the twin and its digital counterpart, resulting in misclassifications.

\section{Architecture \& Results}

\begin{figure*}[h]
\centering
\includegraphics[scale = 0.7]{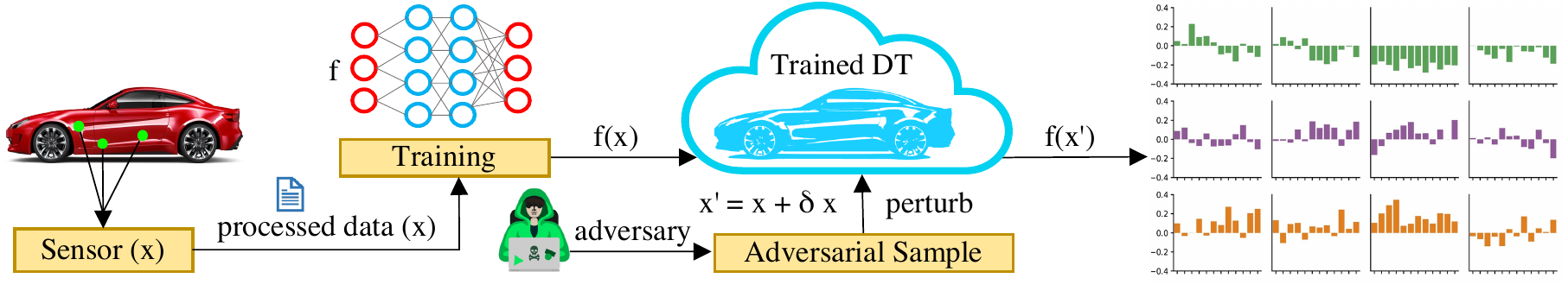}
\caption{Adversarial attack architecture against a trained digital twin.}
\label{figs:architecture}
\vspace{-4mm}
\end{figure*}

In this section, we describe the system's architecture, as shown in Figure \ref{figs:architecture}, as well as the results of our experiments. Vehicles can be viewed as a specialized form of CPS that are outfitted with a variety of sensors that generate a massive amount of operational data. This data is typically collected in a time series format over a period of time and can be used to build intelligent systems such as DT using ML. The physical assets, in our case a vehicle, can then be approximated using DTs to study the behavior of various sensor channels. The advantages of this approach are manifold. For example, a DT model that monitors sensor channel behaviors and detects abnormal patterns early on will not only aid in extending the life cycle and preserve the vehicle's operating performance but will also help to reduce maintenance costs and avoid unplanned downtime. These models, however, are vulnerable to adversarial attacks. In the following paragraphs we describe each component of the attack framework.

In the first phase, the operational data generated by various sensors embedded in vehicular systems are cleaned and processed. The features are chosen in accordance with the subsystem under investigation. As previously stated, DT can be a holistic system or merely be a subset of a system, referred to as subsystems. Our DT is combination of four subsystems: engine, transmission, fuel, and brakes, based on our previous work \cite{neupane2022temporal}. 

\begin{figure}[h]
\vspace{-2mm}
\centering
\includegraphics[width=0.35\textwidth]{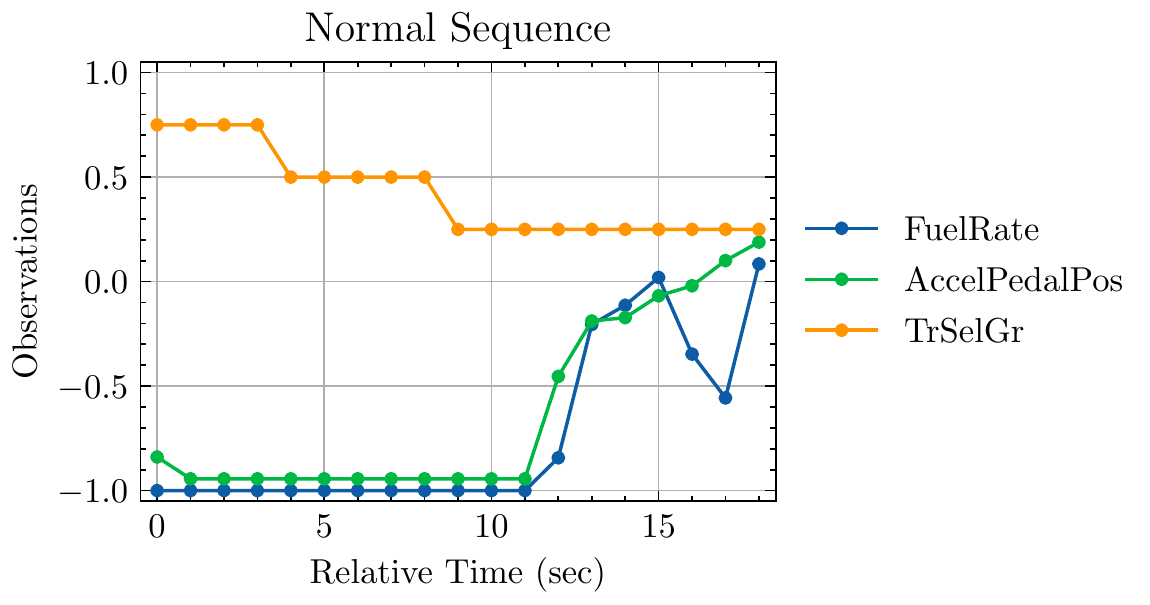}
\vspace{-2mm}
\caption{Normal input sequence. The ADS predicts $NORMAL$ with a Mahalanobis distance of 5.79.}
\label{fig:normal_seq}
\vspace{-4mm}
\end{figure}

A Deep Learning (DL) model is then trained using preprocessed sensor channel data as an input. The resulting DL model $f(x)$ is a trained DT. This model can be used to investigate, monitor, and forecast sensor channel behaviors in a normal scenario, as well as as a predictive maintenance step. Abnormal readings detected in this step can be forwarded to experts for further examination. In the real world, however, adversaries may alter the input data by adding just enough noise to cause the DT model to misclassify a true anomaly as normal. The adversary, for example, can misclassify the output of $f$($x\prime$) by feeding an adversarial input sample ($x\prime$ = x + $\delta x $) to the trained model $f$. An adversarial sample is the small perturbation to the input $x$ used by the function $f$ to predict or forecast that results in misclassified predictions.

To test our model's adversarial robustness, we employ the same evaluation strategies as in our previous work \cite{neupane2022temporal}. In this case, however, we randomly perturb the inputs to the trained model. Given an input sequence, the standard deviation ($\sigma$) is calculated for each channel. The perturbed input sequence is generated by adding Gaussian white noise to the input sequence. The Gaussian white noise is determined by sampling from a normal distribution centered at zero, $\mathcal{N}(0,\,\sigma^{2})$.

The robustness of the anomaly detection system (ADS) in our previous work \cite{neupane2022temporal} is evaluated against test sequences perturbed with white noise. Test sequences contain observations from all 15 channels but only those from Scenario 1 are perturbed (i.e., $FuelRate$, $AccelPetalPos$, $TrSelGr$) for our experiment.



\begin{figure}[h]
\vspace{-2mm}

\centering
\includegraphics[width=0.35\textwidth]{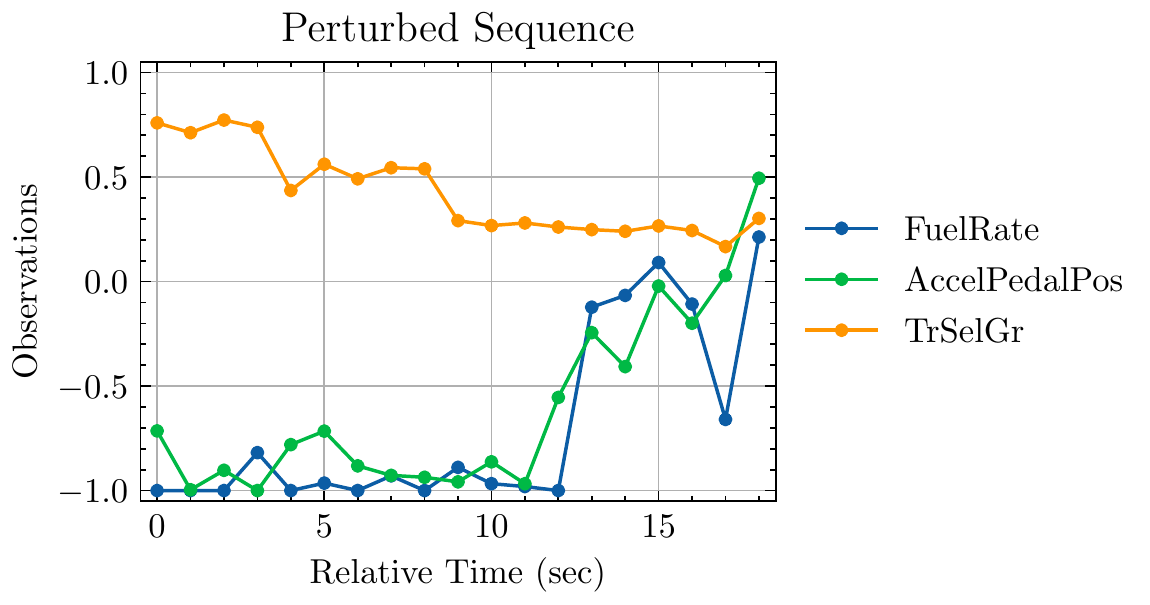}
\vspace{-2mm} 
\caption{Perturbed input sequence. The ADS predicts $ANOMALY$ with a Mahalanobis distance of 9.72.}
\label{fig:perturbed_seq}
\vspace{-4mm}
\end{figure}

\section{Conclusion \& Future Work}
Data-driven Digital Twins are an important concept in the automotive industry because of their ability to classify whether a failure will happen or predict anomalies by learning sensor channel behaviors. However, the robustness of these models against adversarial input samples should be thoroughly tested. In this paper, we demonstrated the vulnerability of these systems and how they can be targeted by an adversary in a white-box attack scenario. As a proof of concept, we attack the trained DT model with an adversarial sample and demonstrate how easily the classifier can be tricked into misclassifying a true anomaly as normal. In the future, we would like to use more sophisticated attack methods such as FSGM and use ART to further validate the robustness of our system.

\section*{Acknowledgement}
Supported by National Science Foundation grant (\#1565484) and by PATENT Lab (Predictive Analytics and TEchnology iNTegration Laboratory) at the Department of Computer Science and Engineering, Mississippi State University.

\bibliographystyle{unsrt}
\bibliography{refs}

\end{document}